\documentclass[twocolumn,showpacs,aps,preprintnumbers,amsmath,amssymb,superscriptaddress,footinbib]{revtex4}

\usepackage{graphicx,subfigure,multirow}
 \usepackage{longtable}
\usepackage{dcolumn}
\usepackage{bm}
\usepackage[all]{xy}
\usepackage{color}
\usepackage[usenames,dvipsnames]{xcolor}
\usepackage[unicode=true,pdfusetitle,
 bookmarks=true,bookmarksnumbered=false,bookmarksopen=false,citecolor=Turquoise,
 breaklinks=false,pdfborder={0 0 1},backref=false,colorlinks=true,pdfpagemode=FullScreen]
 {hyperref}

\newcommand{\GeV}{\textrm{ GeV}} 
\makeatletter

\newcommand{\fmslash}[2][0mu]{%
  \mathchoice
    {\fmsl@sh\displaystyle{#1}{#2}}%
    {\fmsl@sh\textstyle{#1}{#2}}%
    {\fmsl@sh\scriptstyle{#1}{#2}}%
    {\fmsl@sh\scriptscriptstyle{#1}{#2}}}
\newcommand{\fmsl@sh}[3]{%
  \m@th\ooalign{$\hfil#1\mkern#2/\hfil$\crcr$#1#3$}}
\makeatother

\newcommand{\lsim}{{\;\raise0.3ex\hbox{$<$\kern-0.75em\raise-1.1ex\hbox{$\sim$}}\;}}
\newcommand{\gsim}{{\;\raise0.3ex\hbox{$>$\kern-0.75em\raise-1.1ex\hbox{$\sim$}}\;}}

\newcommand{\beq}{\begin{equation}}
\newcommand{\eeq}{\end{equation}}
\newcommand{\bea}{\begin{eqnarray}}
\newcommand{\eea}{\end{eqnarray}}
\mathchardef\minus="002D

\newcommand{\mptvec}{{\vec{\fmslash P}_T}}

\newcommand{\met}{{\fmslash E_T}}
\usepackage{color}

\begin{document}
\title{How to prove that the LHC did not discover dark matter}

\author{Doojin Kim}
\affiliation{Theoretical Physics Department, CERN, CH-1211 Geneva 23, Switzerland}
\author{Konstantin T. Matchev}
\affiliation{Physics Department, University of Florida, Gainesville, FL 32611, USA}
\preprint{CERN-TH-2017-279}

\begin{abstract}
If the LHC is able to produce dark matter particles, they would appear at the end of cascade decay chains, manifesting themselves as missing transverse energy. However, such ``dark matter candidates" may decay invisibly later on. We propose to test for this possibility by studying the effect of particle widths on the observable invariant mass distributions of the visible particles seen in the detector. We consider the simplest non-trivial case of a two-step two-body cascade decay and derive analytically the shapes of the invariant mass distributions, for generic values of the widths of the new particles. We demonstrate that the resulting distortion in the shape of the invariant mass distribution can be significant enough to measure the width of the dark matter ``candidate", ruling it out as the source of the cosmological dark matter. 
\end{abstract}
 

\maketitle

\paragraph*{{\bf Introduction.}} 
$\met$ events at the Large Hadron Collider (LHC) at CERN are motivated by the dark matter problem --- the dark matter particles are stable and weakly interacting, 
and, once produced in the LHC collisions, will escape without leaving a trace inside the detector. This will lead to an imbalance in the transverse momentum of the event, 
known as ``missing transverse momentum" $\mptvec$.\footnote{For historical reasons, the magnitude of this quantity 
is known as the missing transverse energy $\met\equiv |\mptvec|$.} However, the reverse statement is not so obvious --- if we observe an excess of $\met$ 
events at the LHC, how can one be sure that what we are seeing is indeed {\bf the} cosmological dark matter?

The question of {\bf proving} that a $\met$ signal observed at the LHC is indeed due to dark matter, has attracted a lot of attention in the past
\cite{Battaglia:2004mp,Allanach:2004xn,Bourjaily:2005ax,Moroi:2005nc,Birkedal:2005jq,Nojiri:2005ph,Baltz:2006fm,Chung:2007cn}.
The basic idea was to test whether the newly discovered weakly interacting massive particle (WIMP) was consistent with being a thermal relic or not.
The general approach was to assume a specific model, most often some version of low-energy supersymmetry, and then 
attempt to measure all relevant model parameters affecting the thermal relic density calculation. Unfortunately, such an approach is model-dependent;
applies only to thermal relics (for alternative non-thermal scenarios, see \cite{Feng:2003uy,Baer:2014eja});
requires full understanding of the early cosmology; and typically demands a large number of additional measurements, possibly at future (or futuristic) facilities.

Given that {\bf proving} the discovery of dark matter at the LHC is such a difficult task, perhaps one should focus on the opposite
question --- how to {\bf disprove} that the newly found invisible particle is the cosmological dark matter. One possibility is to perform a
precise measurement of its mass, and if the mass is consistent with zero, it may just be one of the Standard Model (SM) neutrinos
instead of a brand new particle \cite{Chang:2009dh}. However, this logic is not ironclad either --- there exist examples where the 
dark matter particles are very light \cite{Gunion:2005rw,Dreiner:2009ic} and cannot be ruled out just on the basis of their small mass.

A much more direct approach would be to test whether the particle which is the source of the $\met$ is indeed stable --- after all, 
we only know that it did not decay inside the detector. If its lifetime is relatively short, so that it does decay outside, but not too far from the detector,
one could attempt to build a dedicated experiment to record such delayed decays. 
In the past, there were proposals to place such supplementary detectors near the D0 experiment at Fermilab 
\cite{Chen:1998awa} and near the LHC \cite{Maki:1997ih}, and these ideas were recently revived in \cite{Chou:2016lxi}.
However, any such experiment is doomed if the dark matter candidate decays invisibly,
e.g., to hidden sector particles \cite{Strassler:2006qa}. 

In this letter we address the worst case scenario, when the dark matter candidate produced at the LHC is unstable and decays invisibly sufficiently quickly.
\begin{figure}[t]
\centering
\includegraphics[width=7.0cm]{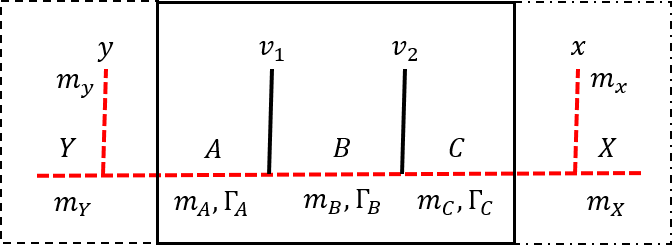} 
\caption{\label{fig:decaychain} The new physics decay chain under study. }
\end{figure} 
For concreteness, we consider the standard new physics decay chain shown inside the solid box of Fig.~\ref{fig:decaychain}:
\bea
A \rightarrow v_1 B \rightarrow v_1 v_2 C\, ,
\label{eq:process}
\eea 
where $v_{1,2}$ are massless SM particles, while $A$, $B$, and $C$ are new particles, with $C$ being the dark matter candidate. 
The canonical example for the processes (\ref{eq:process}) is the neutralino decay 
$\tilde{\chi}_2^0 \rightarrow \ell \tilde{\ell}^\ast \rightarrow \ell\bar{\ell}\tilde{\chi}_1^0$ in supersymmetry \cite{Hinchliffe:1996iu}, 
where $\tilde{\chi}_2^0$ ($\tilde{\chi}_1^0$) is the second-lightest (lightest) neutralino, 
$\tilde\ell$ ($\tilde\ell^\ast$) is a charged (anti-)slepton and $\ell$ ($\bar\ell$) is a SM (anti-)lepton. 
The masses of the particles $A$, $B$ and $C$ are denoted with $m_A$, $m_B$ and $m_C$,
respectively, and in general all three particles will have corresponding widths $\Gamma_A$, $\Gamma_B$ and $\Gamma_C$.
In particular, we shall pay special attention to the case when the dark matter ``candidate" 
$C$ is unstable and thus its decay width $\Gamma_C$ is strictly non-zero.
Our key idea here is to attempt a {\bf direct} measurement of the new particle widths (including $\Gamma_C$)
from the kinematic distributions of the visible decay products $v_1$ and $v_2$.
If one could unambiguously establish experimentally that $\Gamma_C>0$, then $C$ will be ruled out as a dark matter candidate.
Therefore, our first goal is to derive the effect of non-zero widths on the observable kinematics.

\paragraph*{{\bf Pure on-shell case.}} In what follows, we shall be investigating the distribution of the invariant mass $m\equiv m_{v_1v_2}$ of the
two visible particles $v_1$ and $v_2$.
In the purely on-shell case, where all three particles $A$, $B$ and $C$ are exactly on-shell, 
the unit-normalized distribution $dN/dm$ has the well-known ``triangular" shape
\beq
\frac{dN}{dm}=\frac{m}{128\pi^2 m_A^3m_B\Gamma_B}\,,
\label{dNdmon}
\eeq
which extends up to the kinematic endpoint $m_{\text{on}}^{\max}$ 
\beq
m_{\text{on}}^{\max} (m_A,m_B,m_C) \equiv \sqrt{(m_A^2-m_B^2)(m_B^2-m_C^2)}/m_B\,.
\label{monmax}
\eeq
The validity of (\ref{dNdmon}) is ensured (at tree-level) as long as the narrow width approximation holds and there are no significant polarization effects. 
We shall now investigate how the result (\ref{dNdmon}) is modified in the case of non-negligible widths $\Gamma_A$,
$\Gamma_B$ and, most importantly, $\Gamma_C$. For simplicity, we shall be turning on those widths one at a time.

\paragraph*{{\bf Non-negligible $\Gamma_B$.}} 

As a warm-up, we begin with the case when only $B$ is relatively broad, $\Gamma_B\ne 0$. In that case, the narrow-width result 
(\ref{dNdmon}) gets modified to \cite{Grossman:2011nh}
\beq
\frac{dN}{dm}=\frac{m}{128 \pi^3 m_A^3} \int_{s_-}^{s_+} \frac{ds}{(s-m_B^2)^2+m_B^2\Gamma_B^2},
\label{dNdmoffB}
\eeq
where
\beq
s_{\pm}\equiv \frac{1}{2} \left[m_A^2+m_C^2-m^2\pm\lambda^{1/2}(m_A^2,m_C^2,m^2)\right]\, ,
\label{spmB}
\eeq
and $\lambda(x,y,z)\equiv x^2+y^2+z^2-2xy-2yz-2xz$. In the limit of massless $v_1$ and $v_2$,
the lower endpoint of (\ref{dNdmoffB}) is at $m=0$, while the upper endpoint, $m_{\Gamma_B}^{\max}$, is obtained by solving the equation $s_- = s_+$, which results in
\beq
m_{\Gamma_B}^{\max} = m_A -m_B\,,
\label{mmaxGB}
\eeq
a result identical to the one for the direct three-body decay 
\beq
A\to v_1 v_2 C.
\label{3bodyAtoC}
\eeq  

Note that in the narrow width approximation limit of $\Gamma_B/m_B \rightarrow 0$, the integrand in (\ref{dNdmoffB})  becomes 
\beq
\underset{\frac{\Gamma_B}{m_B} \rightarrow 0}{\lim }\frac{1}{(s-m_B^2)^2+m_B^2\Gamma_B^2} =
\frac{\pi}{m_B^3\Gamma_B} \delta\left(\frac{s}{m_B^2}-1 \right)
\eeq
and we recover the purely on-shell result (\ref{dNdmon}).

\begin{figure}[tbp]
\centering
\includegraphics[width=4.2cm]{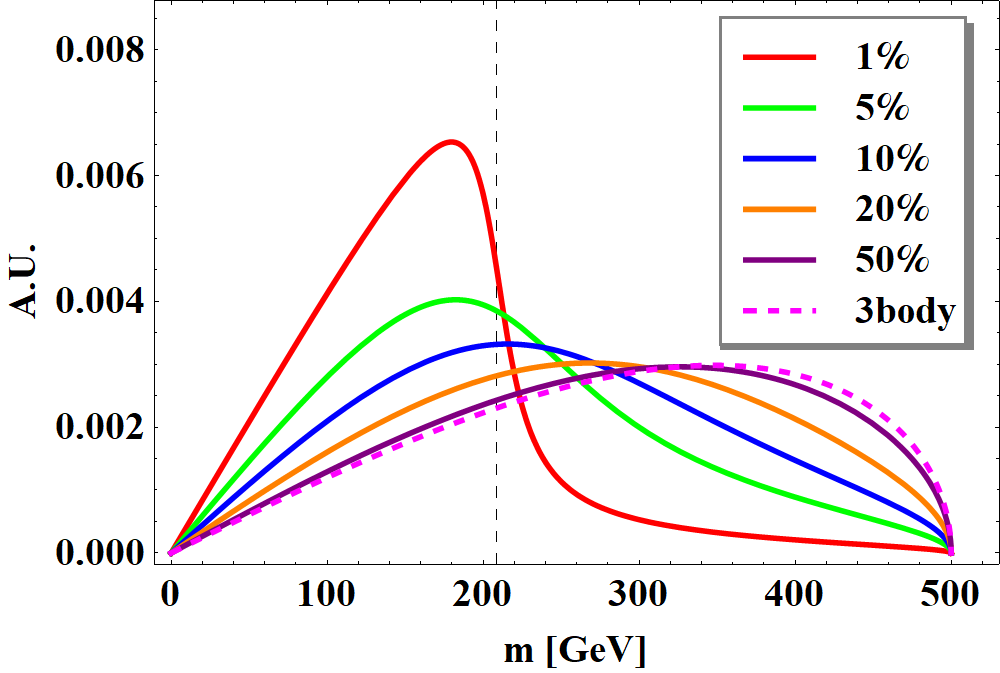} 
\includegraphics[width=4.2cm]{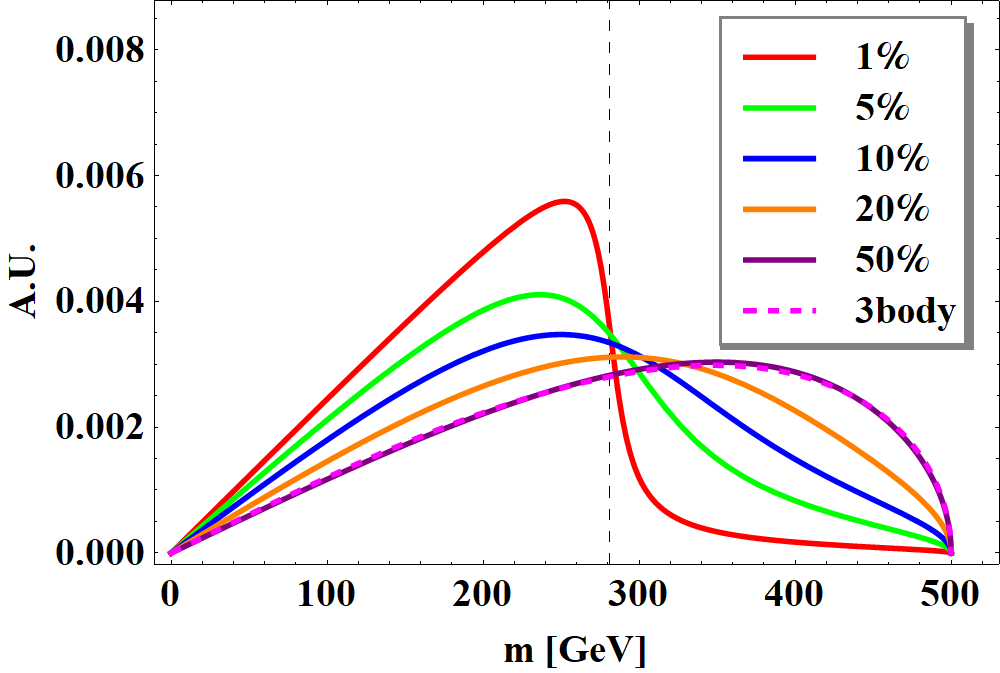} 
\caption{\label{fig:gammaB} The solid lines represent unit-normalized
invariant mass distributions (\ref{dNdmoffB}) for 
$(m_A,m_B,m_C)=(1000,970,500)$ GeV (left panel)
and 
$(m_A,m_B,m_C)=(1000,530,500)$ GeV (right panel),
with negligible $\Gamma_A$ and $\Gamma_C$ and several 
different choices of $\Gamma_B/m_B$ as shown in the legends.
The magenta dashed curve corresponds to the case of a pure three-body decay (e.g., $m_B\gg m_A$).  }
\end{figure}

Fig.~\ref{fig:gammaB} illustrates the effect of a finite width $\Gamma_B$ on the invariant mass distribution (\ref{dNdmoffB}). 
In general, one should expect sizable effects whenever the width $\Gamma_B$ is comparable to a relevant mass 
splitting,\footnote{This point was overlooked in the previous analysis of Ref.~\cite{Grossman:2011nh}.}
e.g., $m_A-m_B$ (left panel) or $m_B-m_C$ (right panel).
The solid lines depict the invariant mass distribution (\ref{dNdmoffB}) for several different values of $\Gamma_B/m_B$, from $1\%$ (red lines) all the way to $50\%$
(purple lines).
For comparison, the $m$ distribution for the three-body decay (\ref{3bodyAtoC}) is shown by the magenta dashed curve. 
We see that initially, as the width $\Gamma_B$ is relatively small, the shape of the distribution still resembles the triangular shape of (\ref{dNdmon}),
but there are a certain number of events which leak out beyond the nominal upper kinematic endpoint (\ref{monmax}).
As the width $\Gamma_B$ increases, so does the fraction of events which leak out, and very soon, for $\Gamma_B/m_B \sim 5-10\%$,
no discernible endpoint is visible at all at the location (vertical dashed line) predicted by (\ref{monmax}). Instead, we obtain a relatively broad distribution 
which terminates at the new kinematic endpoint (\ref{mmaxGB}). Eventually, as the width $\Gamma_B$ further increases,
the distribution asymptotes to the magenta dashed line corresponding to the case of the three-body decay (\ref{3bodyAtoC}).

Fig.~\ref{fig:gammaB} demonstrates that the effect of a finite $\Gamma_B$ on the invariant mass distribution 
(\ref{dNdmoffB}) can be quite significant --- for one, all curves in the figure have shapes which are clearly different from the 
triangular shape (\ref{dNdmon}) obtained in the limit of $\Gamma_B=0$. At the same time, unless the $B$ resonance is extremely broad ($\Gamma_B\sim m_B$),
the obtained distribution is also distinguishable from that of a three-body decay (\ref{3bodyAtoC}).
We thus conclude that the observation of a non-trivial invariant mass shape like the ones seen in Fig.~\ref{fig:gammaB}
would not only suggest a finite value for $\Gamma_B$, but will also allow its measurement with a decent precision.

\begin{figure}[tbp]
\centering
\includegraphics[height=4.0cm]{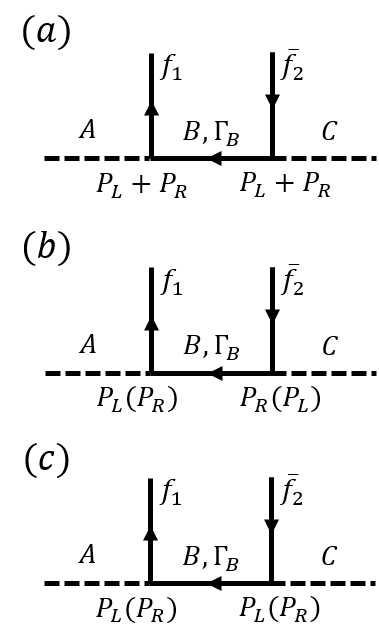} 
\includegraphics[height=4.0cm]{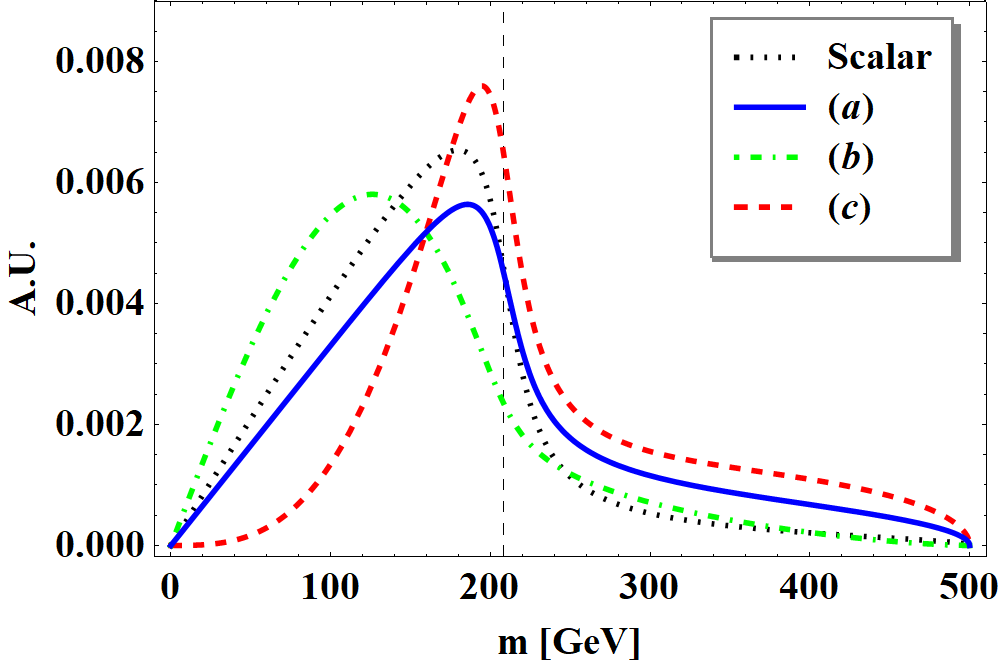} 
\caption{\label{fig:mBvarying} Left panel: Three different fermion chirality structures for the boxed decay chain of Fig.~\ref{fig:decaychain}: 
(a) vectorlike couplings, (b) opposite chiralities, and (c) same chiralities at the neighboring fermion vertices.
Right panel: Unit-normalized invariant mass distributions for those three cases, compared to the pure scalar theory result (\ref{dNdmoffB}) (black dotted line), for 
$(m_A,m_B,m_C)=(1000,970,500)$ GeV and $\Gamma_B/m_B=1\%$.}
\end{figure}

Before we move on to the case of a non-negligible $\Gamma_C$, let us briefly comment on the effect of spin correlations.
Our previous results were obtained in the pure phase space limit, where the width dependence comes only from the $B$ propagator.
However, these results would be valid only if all involved particles are spin 0, which is unrealistic --- the SM particles $v_1$ and $v_2$ are
fermions (leptons or quark-initiated jets). Therefore, some non-trivial chiralities are present in the interaction vertices,
as shown in the left panel of Fig.~\ref{fig:mBvarying}, where for concreteness we have chosen the intermediate particle $B$ to be a fermion.\footnote{The alternative choice 
is for $B$ to be a boson, while $A$ and $C$ are fermions.}
In general, the fermion couplings are arbitrary mixtures of left-handed and right-handed chiral couplings proportional to 
$P_L\equiv (1-\gamma_5)/2$ and $P_R\equiv (1+\gamma_5)/2$, respectively. In Fig.~\ref{fig:mBvarying}, we contrast three
special cases: (a) vectorlike couplings, (b) opposite chiralities at the two vertices and (c) the same chiralities at the two vertices.
Then, the spin-averaged matrix element squared receives an additional contribution proportional to 
\beq
\overline{|{\cal M}|^2 }\sim 
\left\{
\begin{array}{ll}
(m_A^2-s)(s-m_C^2)-m^2s, \text{for Fig.~\ref{fig:mBvarying}(b)}\\ [2mm]
m^2\left(\frac{\Gamma_B^2}{4}+m_B^2\right), \text{for Fig.~\ref{fig:mBvarying}(c)}.
\end{array}
\right.
\label{Msqbc}
\eeq
Therefore, the result for vectorlike couplings (Fig.~\ref{fig:mBvarying}(a)) is simply the sum of these two cases
(times a factor of 2 due to $L\leftrightarrow R$ exchange)
\beq
\overline{|{\cal M}|^2 }\sim 
2(m_A^2-s)(s-m_C^2)+2m^2\left(\frac{\Gamma_B^2}{4}+m_B^2-s\right).
\label{Msqa}
\eeq
The chirality effects (\ref{Msqbc},\ref{Msqa}) on the shape of the invariant mass distribution are illustrated
in the right panel of Fig.~\ref{fig:mBvarying}, for a mass spectrum  $(m_A,m_B,m_C)=(1000,970,500)$ GeV and 
$\Gamma_B/m_B=1\%$. For reference, the black dotted line shows the pure scalar theory result (\ref{dNdmoffB}).
The green dot-dashed and the red dashed lines represent the distributions obtained in the presence of spin correlations as in 
Fig.~\ref{fig:mBvarying}(b) and Fig.~\ref{fig:mBvarying}(c), respectively. The case of vectorlike couplings, Fig.~\ref{fig:mBvarying}(a),
is then obtained by simply adding those two distributions (blue solid line). 
In the narrow width approximation, for vectorlike couplings one would recover the phase space result (\ref{dNdmon}), 
since the spin correlations from Fig.~\ref{fig:mBvarying}(b) and Fig.~\ref{fig:mBvarying}(c) would cancel exactly.
However, in the presence of non-trivial width effects as in (\ref{Msqbc}), the cancellation is incomplete and 
even the case of vector-like couplings is markedly different from the pure scalar theory result 
(compare the blue solid and black dotted lines in Fig.~\ref{fig:mBvarying}) \cite{Wang:2006hk}.

\paragraph*{\bf Non-negligible $\Gamma_C$.} 
We now consider perhaps the most interesting case, when the dark matter 
candidate (particle $C$) has a non-vanishing width, $\Gamma_C \neq 0$, due to
an invisible decay to two dark sector particles $X$ and $x$, as shown in the right (dot-dashed) boxed extension of Fig.~\ref{fig:decaychain}.
Under those circumstances, we find that the shape of the invariant mass distribution is given by
\beq
\frac{dN}{dm}=\frac{m}{2048\pi^4 m_A^3m_B\Gamma_B} \int_{s_-}^{s_+}\frac{ds}{s} \frac{\lambda^{1/2}(s,m_X^2,m_x^2)}{(s-m_C^2)^2+m_C^2\Gamma_C^2}\, ,
\label{dNdmoffC}
\eeq
where $m_X$ and $m_x$ are the respective masses of the hidden sector particles $X$ and $x$ and
\bea
s_- \equiv (m_X+m_x)^2, ~~~ s_+ \equiv m_B^2\left(1-\frac{m^2}{m_A^2-m_B^2}\right)\,.
\label{spmC}
\eea
As before, the upper kinematic endpoint, $m_{\Gamma_C}^{\max}$, of the distribution (\ref{dNdmoffC})
is found from $s_-=s_+$, which yields
\bea
m_{\Gamma_C}^{\max} =\sqrt{(m_A^2-m_B^2)\lbrace m_B^2-(m_X+m_x)^2 \rbrace }/m_B\,.
\eea
Comparing to (\ref{monmax}), we notice that  
\beq
m_{\Gamma_C}^{\max} = m_{\text{on}}^{\max} (m_A,m_B,m_X+m_x),
\label{mGCmaxrelation}
\eeq
which is easily understood as the limit when $C$ becomes extremely off-shell. 

\begin{figure}[tbp]
\centering
\includegraphics[width=5.5cm]{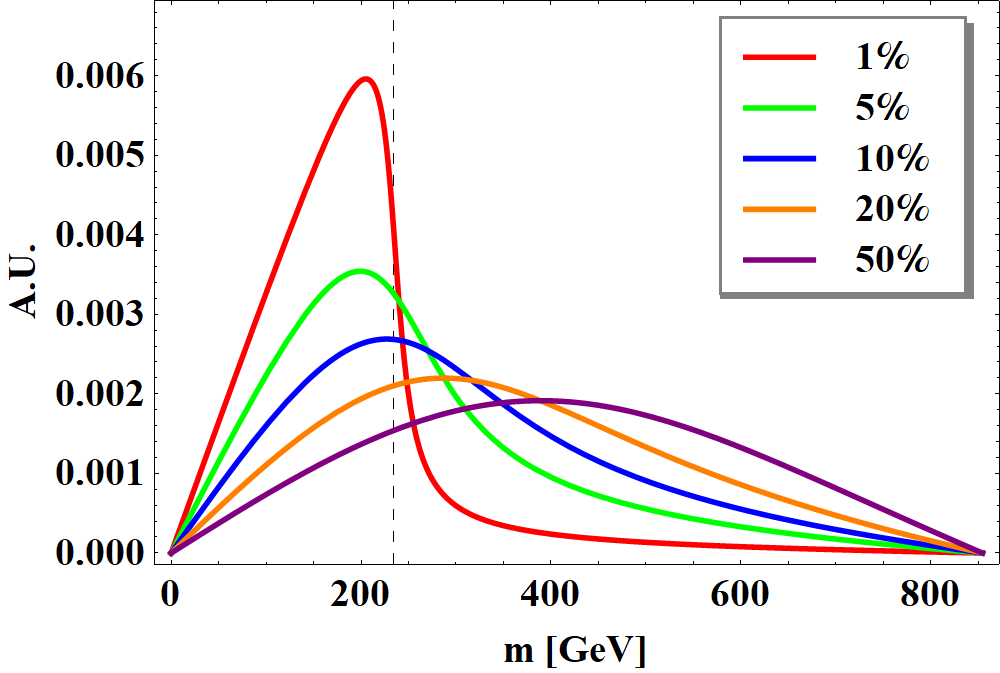} 
\caption{\label{fig:gammavarying} 
Unit-normalized invariant mass distributions for $(m_A,m_B,m_C)=(1000,520,500)$ GeV and several different values of  $\Gamma_C/m_C$ as shown in the legend. 
We assume that particle $C$ further decays invisibly to two massless particles $X$ and $x$,  $C\to Xx$, as shown in the dot-dashed box of Fig.~\ref{fig:decaychain}. 
}
\end{figure} 

In analogy to Fig.~\ref{fig:mBvarying}, Fig.~\ref{fig:gammavarying} illustrates the impact of 
the non-vanishing width $\Gamma_C$ on the shape of the invariant mass distribution (\ref{dNdmoffC}). 
We take the mass spectrum to be $(m_A,m_B,m_C)=(1000,520,500)$ GeV and 
again vary the dimensionless ratio $\Gamma_C/m_C$ from 1\% to 50\% as indicated in the legend.
For concreteness, we assume the hidden sector particles $X$ and $x$ to be massless, i.e., $m_X=m_x=0$,
in which case the distributions in Fig.~\ref{fig:gammavarying}  have a common upper kinematic endpoint $m_{\Gamma_C}^{\max}=\sqrt{m_A^2-m_B^2}=854 \GeV$.

Fig.~\ref{fig:gammavarying} demonstrates that the effect of $\Gamma_C$ can be quite drastic. 
Even when the width $\Gamma_C$ is as small as 1\% of the resonance mass $m_C$, the 
shape of the distribution is visibly distorted from the standard triangular shape (\ref{dNdmon}), 
and a sizable fraction of events are already leaking out beyond the expected kinematic endpoint (\ref{monmax}),
which is indicated with the vertical dashed line. Increasing the width to $\Gamma_C\sim 0.05\, m_C$ 
appears already sufficient to render the triangular shape unrecognizable and indicate the presence of off-shell effects.

\paragraph*{\bf Non-negligible $\Gamma_A$.} Finally, for completeness we also consider the case where the decay
width of particle $A$ is non-negligible, $\Gamma_A \neq 0$. This case is a little bit more model-dependent, since
we must know how to sample the 4-momentum squared, $p_A^2$, of particle $A$. 
One simple possibility is that $A$ is the decay product of a narrow resonance $Y$ with mass $m_Y$, 
$Y\to yA$, as shown in the left (dashed) boxed extension of Fig.~\ref{fig:decaychain}.
Under those circumstances, the invariant mass distribution is given by
\beq
\frac{dN}{dm} = \frac{m}{2048 \pi^4 m_Y^3m_B\Gamma_B}\int_{s_-}^{s_+}\frac{ds}{s} \frac{\lambda^{1/2}(m_Y^2,m_y^2,s)}{(s-m_A^2)^2+m_A^2\Gamma_A^2} \,,
\label{dNdmoffA}
\eeq
where $m_Y$ and $m_y$ are the masses of the particles $Y$ and $y$, respectively, while
\bea
s_-\equiv m_B^2\left(1+\frac{m^2}{m_B^2-m_C^2}\right),~~~ s_+\equiv (m_Y-m_y)^2\,.
\label{spmA}
\eea
The upper kinematic endpoint, $m_{\Gamma_A}^{\max}$, of the distribution (\ref{dNdmoffA})
is again found from $s_-=s_+$:
\bea
m_{\Gamma_A}^{\max}=\sqrt{ \lbrace (m_Y-m_y)^2-m_B^2)(m_B^2-m_C)^2 }/m_B\,,
\eea
and can be equivalently interpreted as
\beq
m_{\Gamma_A}^{\max}= m_{\text{on}}^{\max} (m_Y-m_y,m_B,m_C).
\label{mGAmaxrelation}
\eeq

\begin{figure}[tbp]
\centering
\includegraphics[width=5.5cm]{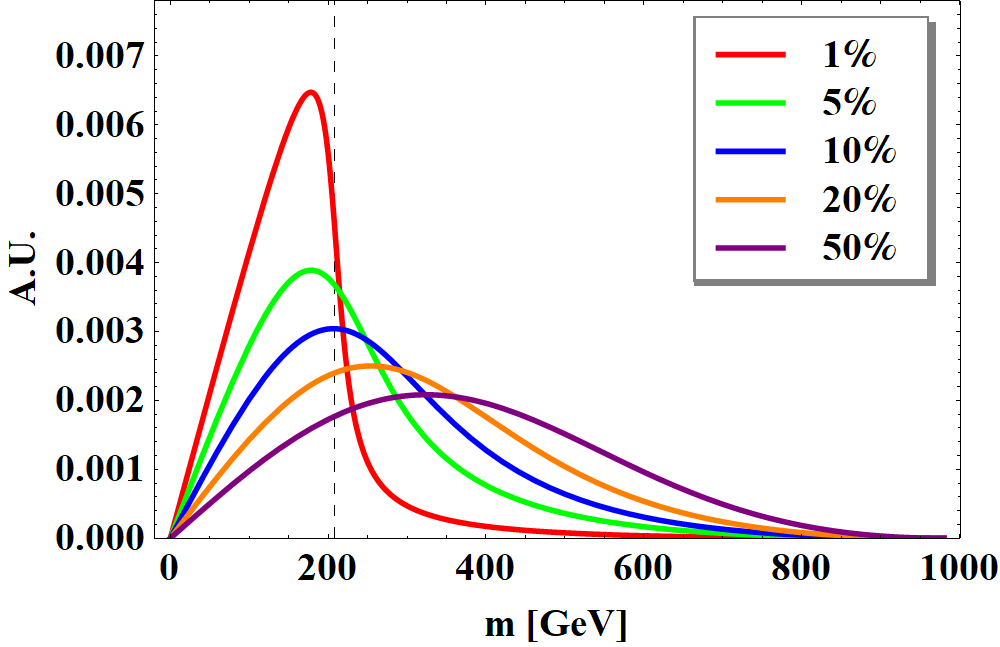}
\caption{\label{fig:gammaAvarying} Unit-normalized invariant mass distributions for $(m_Y,m_A,m_B,m_C)=(1500,1000,970,500)$ GeV and several different values of  $\Gamma_A/m_A$ as shown in the legend. We assume that $A$ results from the decay of a parent particle $Y$, $Y\to yA$ (see the dashed box of Fig.~\ref{fig:decaychain}).
The particle $y$ is assumed massless, and may or may not be visible in the detector. 
}
\end{figure} 

Fig.~\ref{fig:gammaAvarying} shows the effect of a non-vanishing width $\Gamma_A$ on the shape of the
invariant mass distribution (\ref{dNdmoffA}). The mass spectrum is chosen as
$(m_Y,m_A,m_B,m_C)=(1500,1000,970,500)$ GeV 
and the dimensionless ratio $\Gamma_A/m_A$ is again varied from 1\% to 50\%,
as indicated in the legend. For concreteness, we assume that the additional final state particle $y$ is massless,
then all distributions in Fig.~\ref{fig:gammaAvarying} have a common kinematic endpoint
$m_{\text{on}}^{\max} (m_Y,m_B,m_C)=980 \GeV$, as
predicted by (\ref{mGAmaxrelation}). Once again, we observe that
even a width of only $1\%$ leads to a noticeable change in the expected triangular shape 
and an overflow of events beyond the nominal 
kinematic endpoint of 208.3 GeV predicted by (\ref{monmax}) and denoted by the vertical dashed line. 
As the width is further increased, the shape distortion becomes quite significant, 
confirming the sensitivity to the value of $\Gamma_A$.

\paragraph*{{\bf Summary and outlook.}} 
We derived the  effects of non-zero particle widths on the observable invariant mass distribution $dN/dm$ in the case of the decay chain of Fig.~\ref{fig:decaychain}.
We showed that the shape of the distribution can be very sensitive to the widths and therefore can be used to perform a measurement of $\Gamma_A$,
$\Gamma_B$ and, most importantly, $\Gamma_C$, thus directly probing the nature of the dark matter candidate $C$, 
which appears invisible in the detector. Our results for these three cases can be compactly summarized as
\bea
\frac{dN}{dm} \sim m\int_{s_{i-}}^{s_{i+}}ds\,\frac{1}{(s-m_i^2)^2+m_i^2\Gamma_i^2}\, F_i(s)  \,,
\eea
where $i=\{A, B, C\}$, the integration limits $s_{i\pm}$ are given by eqs.~(\ref{spmA}), (\ref{spmB}) and (\ref{spmC}), respectively, while
\bea
F_i(s)=\left\{
\begin{array}{l l}
\frac{\lambda^{1/2}(m_{Y}^2,m_{y}^2,s)}{s}, & \hbox{for }i=A;  \\ [2mm]
1, & \hbox{for }i=B; \\ [2mm]
\frac{\lambda^{1/2}(s,m_{X}^2,m_{x}^2)}{s}, & \hbox{for }i=C. 
\end{array}\right. 
\eea

One should be mindful of the fact that there are other factors which also affect the shape of the invariant mass distribution
$dN/dm$. On the theoretical side, there could be spin correlations \cite{Barr:2004ze,Smillie:2005ar,Wang:2006hk,Burns:2008cp}, interference \cite{Birkedal:2005cm,Fuchs:2014ola}
and higher order effects \cite{Drees:2006um,Beneke:2016igc}.
On the experimental side, the cuts and the detector resolution will also play a role in this measurement.
However, these effects are well known and under control, and can be readily accounted for 
(see, e.g., the kinematic endpoint measurements in \cite{Chatrchyan:2013boa}).
Furthermore, the width measurement relies mostly on the events {\bf above} the nominal kinematic endpoint
(\ref{monmax}), while all those effects impact mostly the softer part of the distribution $dN/dm$.
We are therefore optimistic that such width measurements will be feasible, once a 
sufficiently strong and clean missing energy signal of new physics is observed at the LHC. 

\section*{Acknowledgments}
\vspace{-0.2cm}
We would like to thank Gennaro Corcella and Rakhi Mahbubani for insightful discussions. 
This work is  supported in part by a US Department of Energy grant DE-SC0010296.  DK was supported in part by the LHC Theory Initiative postdoctoral fellowship (NSF Grant No. PHY-0969510), and presently supported by the Korean Research Foundation (KRF) through the CERN-Korea Fellowship program.




\begin{thebibliography}{999}

\bibitem{Battaglia:2004mp} 
  M.~Battaglia, I.~Hinchliffe and D.~Tovey,
  ``Cold dark matter and the LHC,''
  J.\ Phys.\ G {\bf 30}, R217 (2004)
  doi:10.1088/0954-3899/30/10/R01
  [hep-ph/0406147].

\bibitem{Allanach:2004xn} 
  B.~C.~Allanach, G.~Belanger, F.~Boudjema and A.~Pukhov,
  ``Requirements on collider data to match the precision of wmap on supersymmetric dark matter,''
  JHEP {\bf 0412}, 020 (2004)
  doi:10.1088/1126-6708/2004/12/020
  [hep-ph/0410091].
    
\bibitem{Bourjaily:2005ax} 
  J.~L.~Bourjaily and G.~L.~Kane,
  ``What is the cosmological significance of a discovery of wimps at colliders or in direct experiments?,''
  hep-ph/0501262.

\bibitem{Moroi:2005nc} 
  T.~Moroi, Y.~Shimizu and A.~Yotsuyanagi,
  ``Reconstructing dark matter density with $e^+e^-$ linear collider in focus-point supersymmetry,''
  Phys.\ Lett.\ B {\bf 625}, 79 (2005)
  doi:10.1016/j.physletb.2005.07.068
  [hep-ph/0505252].
     
\bibitem{Birkedal:2005jq} 
  A.~Birkedal {\it et al.},
  ``Testing cosmology at the ILC,''
  eConf C {\bf 050318}, 0708 (2005)
  [hep-ph/0507214].

\bibitem{Nojiri:2005ph} 
  M.~M.~Nojiri, G.~Polesello and D.~R.~Tovey,
  ``Constraining dark matter in the MSSM at the LHC,''
  JHEP {\bf 0603}, 063 (2006)
  doi:10.1088/1126-6708/2006/03/063
  [hep-ph/0512204].
    
 \bibitem{Baltz:2006fm} 
  E.~A.~Baltz, M.~Battaglia, M.~E.~Peskin and T.~Wizansky,
  ``Determination of dark matter properties at high-energy colliders,''
  Phys.\ Rev.\ D {\bf 74}, 103521 (2006)
  doi:10.1103/PhysRevD.74.103521
  [hep-ph/0602187].
  
\bibitem{Chung:2007cn} 
  D.~Chung, L.~Everett, K.~Kong and K.~T.~Matchev,
  ``Connecting LHC, ILC, and Quintessence,''
  JHEP {\bf 0710}, 016 (2007)
  doi:10.1088/1126-6708/2007/10/016
  [arXiv:0706.2375 [hep-ph]].
    
\bibitem{Feng:2003uy} 
  J.~L.~Feng, A.~Rajaraman and F.~Takayama,
  ``SuperWIMP dark matter signals from the early universe,''
  Phys.\ Rev.\ D {\bf 68}, 063504 (2003)
  doi:10.1103/PhysRevD.68.063504
  [hep-ph/0306024].

\bibitem{Baer:2014eja} 
  H.~Baer, K.~Y.~Choi, J.~E.~Kim and L.~Roszkowski,
  ``Dark matter production in the early Universe: beyond the thermal WIMP paradigm,''
  Phys.\ Rept.\  {\bf 555}, 1 (2015)
  doi:10.1016/j.physrep.2014.10.002
  [arXiv:1407.0017 [hep-ph]].

\bibitem{Chang:2009dh} 
  S.~Chang and A.~de Gouvea,
  ``Neutrino alternatives for missing energy events at colliders,''
  Phys.\ Rev.\ D {\bf 80}, 015008 (2009)
  doi:10.1103/PhysRevD.80.015008
  [arXiv:0901.4796 [hep-ph]].

\bibitem{Gunion:2005rw} 
  J.~F.~Gunion, D.~Hooper and B.~McElrath,
  ``Light neutralino dark matter in the NMSSM,''
  Phys.\ Rev.\ D {\bf 73}, 015011 (2006)
  doi:10.1103/PhysRevD.73.015011
  [hep-ph/0509024].
  
\bibitem{Dreiner:2009ic} 
  H.~K.~Dreiner, S.~Heinemeyer, O.~Kittel, U.~Langenfeld, A.~M.~Weber and G.~Weiglein,
  ``Mass Bounds on a Very Light Neutralino,''
  Eur.\ Phys.\ J.\ C {\bf 62}, 547 (2009)
  doi:10.1140/epjc/s10052-009-1042-y
  [arXiv:0901.3485 [hep-ph]].
        
\bibitem{Chen:1998awa} 
  C.~H.~Chen and J.~F.~Gunion,
  ``Probing gauge mediated supersymmetry breaking models at the Tevatron via delayed decays of the lightest neutralino,''
  Phys.\ Rev.\ D {\bf 58}, 075005 (1998)
  doi:10.1103/PhysRevD.58.075005
  [hep-ph/9802252].
  
\bibitem{Maki:1997ih} 
  K.~Maki and S.~Orito,
  ``Hadron colliders as the 'neutralino factory': Search for a slow decay of the lightest neutralino at the CERN LHC,''
  Phys.\ Rev.\ D {\bf 57}, 554 (1998)
  doi:10.1103/PhysRevD.57.554
  [hep-ph/9706382].
    
\bibitem{Chou:2016lxi} 
  J.~P.~Chou, D.~Curtin and H.~J.~Lubatti,
  ``New Detectors to Explore the Lifetime Frontier,''
  Phys.\ Lett.\ B {\bf 767}, 29 (2017)
  doi:10.1016/j.physletb.2017.01.043
  [arXiv:1606.06298 [hep-ph]].

\bibitem{Strassler:2006qa} 
  M.~J.~Strassler,
  ``Possible effects of a hidden valley on supersymmetric phenomenology,''
  hep-ph/0607160.

\bibitem{Hinchliffe:1996iu} 
  I.~Hinchliffe, F.~E.~Paige, M.~D.~Shapiro, J.~Soderqvist and W.~Yao,
  ``Precision SUSY measurements at CERN LHC,''
  Phys.\ Rev.\ D {\bf 55}, 5520 (1997)
  doi:10.1103/PhysRevD.55.5520
  [hep-ph/9610544].


\bibitem{Grossman:2011nh} 
  Y.~Grossman, M.~Martone and D.~J.~Robinson,
  ``Kinematic Edges with Flavor Oscillation and Non-Zero Widths,''
  JHEP {\bf 1110}, 127 (2011)
  doi:10.1007/JHEP10(2011)127
  [arXiv:1108.5381 [hep-ph]].
 
 \bibitem{Wang:2006hk} 
  L.~T.~Wang and I.~Yavin,
  ``Spin measurements in cascade decays at the LHC,''
  JHEP {\bf 0704}, 032 (2007)
  doi:10.1088/1126-6708/2007/04/032
  [hep-ph/0605296].
    

 \bibitem{Barr:2004ze} 
  A.~J.~Barr,
  ``Determining the spin of supersymmetric particles at the LHC using lepton charge asymmetry,''
  Phys.\ Lett.\ B {\bf 596}, 205 (2004)
  doi:10.1016/j.physletb.2004.06.074
  [hep-ph/0405052].

\bibitem{Smillie:2005ar} 
  J.~M.~Smillie and B.~R.~Webber,
  ``Distinguishing spins in supersymmetric and universal extra dimension models at the large hadron collider,''
  JHEP {\bf 0510}, 069 (2005)
  doi:10.1088/1126-6708/2005/10/069
  [hep-ph/0507170].
  
\bibitem{Burns:2008cp} 
  M.~Burns, K.~Kong, K.~T.~Matchev and M.~Park,
  ``A General Method for Model-Independent Measurements of Particle Spins, Couplings and Mixing Angles in Cascade Decays with Missing Energy at Hadron Colliders,''
  JHEP {\bf 0810}, 081 (2008)
  doi:10.1088/1126-6708/2008/10/081
  [arXiv:0808.2472 [hep-ph]].
      
 \bibitem{Birkedal:2005cm} 
  A.~Birkedal, R.~C.~Group and K.~Matchev,
  ``Slepton mass measurements at the LHC,''
  eConf C {\bf 050318}, 0210 (2005)
  [hep-ph/0507002].
  
  \bibitem{Fuchs:2014ola} 
  E.~Fuchs, S.~Thewes and G.~Weiglein,
  ``Interference effects in BSM processes with a generalised narrow-width approximation,''
  Eur.\ Phys.\ J.\ C {\bf 75}, 254 (2015)
  doi:10.1140/epjc/s10052-015-3472-z
  [arXiv:1411.4652 [hep-ph]].
  
\bibitem{Drees:2006um} 
  M.~Drees, W.~Hollik and Q.~Xu,
  ``One-loop calculations of the decay of the next-to-lightest neutralino in the MSSM,''
  JHEP {\bf 0702}, 032 (2007)
  doi:10.1088/1126-6708/2007/02/032
  [hep-ph/0610267].
    
\bibitem{Beneke:2016igc} 
  M.~Beneke, L.~Jenniches, A.~Mück and M.~Ubiali,
  ``Radiative distortion of kinematic edges in cascade decays,''
  Phys.\ Lett.\ B {\bf 770}, 539 (2017)
  doi:10.1016/j.physletb.2017.04.018
  [arXiv:1611.08166 [hep-ph]].

\bibitem{Chatrchyan:2013boa} 
  S.~Chatrchyan {\it et al.} [CMS Collaboration],
  ``Measurement of masses in the $t \bar{t}$ system by kinematic endpoints in pp collisions at $\sqrt{s}$ = 7 TeV,''
  Eur.\ Phys.\ J.\ C {\bf 73}, 2494 (2013)
  doi:10.1140/epjc/s10052-013-2494-7
  [arXiv:1304.5783 [hep-ex]].
  
  
    
\end{thebibliography}
\end{document}